\documentclass[11pt,oneside,reqno,reprint,onecolumn,nofootinbib]{revtex4-1}
\usepackage{amsfonts}
\usepackage{amsmath}
\usepackage[a4paper,left=1in,right=1in,top=1in,bottom=1in,footskip=.25in]{geometry}
\usepackage{graphicx}
\usepackage{subcaption}
\usepackage[noabbrev]{cleveref}
\usepackage{amsmath}
\usepackage{amssymb}
\usepackage{epstopdf}
\usepackage{color}
\usepackage{hyperref}

\begin{document}

\title{The Shape of Bouncing Universes}
\author{John D. Barrow}
\email[(Corresponding Author)]{J.D.Barrow@damtp.cam.ac.uk}
\author{Chandrima Ganguly}
\email{C.Ganguly@damtp.cam.ac.uk}
\affiliation{DAMTP, Centre for Mathematical Sciences,\\
University of Cambridge,\\
Wilberforce Rd.,\\
Cambridge CB3 0WA\\
United Kingdom}
\date{28th March, 2017}

\begin{abstract}
What happens to the most general closed oscillating universes in general
relativity? We sketch the development of interest in cyclic universes from
the early work of Friedmann and Tolman to modern variations introduced by
the presence of a cosmological constant. Then we show what happens in the
cyclic evolution of the most general closed anisotropic universes provided
by the Mixmaster universe. We show that in the presence of entropy increase
its cycles grow in size and age, increasingly approaching flatness. But
these cycles also grow increasingly anisotropic at their expansion maxima.
If there is a positive cosmological constant, or dark energy, present then
these oscillations always end and the last cycle evolves from an anisotropic
inflexion point towards a de Sitter future of everlasting expansion. \newline
\newline
\newline
\newline
\newline

\textbf{Essay written for the Gravity Research Foundation 2017 Awards}
\end{abstract}

\maketitle

Cyclic universes have a long history of ups and downs. Hurdling myths about
the death and rebirth of a universe, the inception of the general theory of
relativity led quickly to precise possibilities. Alexander Friedmann noticed
the scope for periodically expanding and contracting closed universes in his
first cosmological solutions of Einstein's equations \cite{fried}. The next
important consideration of the physical reality of such eternal, cyclic
universes was made by Richard Tolman in 1928 \cite{tol}. Tolman, a chemist
as well as a physicist, wanted to introduce the thermodynamic arrow of time
into these cosmological models by including the second law of
thermodynamics. He made a striking discovery. The entropy increase from
cycle to cycle in an oscillating closed Friedmann universe increased the
maximum size and timespan of successive cycles (Figure 1). Today, we would
interpret that as driving successive cycles towards `flatness'. Successive
cycles undergo longer and longer periods of their evolution in proximity to
the expansion maxima. The greater the number of past cycles, the closer are
dynamics to spatial flatness. This time asymmetry can be interpreted
physically if the steady entropy increase from cycle to cycle to result from
the transfer of ordered energy, in pressureless dust, to disordered
blackbody radiation. This entropy-increasing transfer of energy from
pressureless matter to high-pressure radiation creates an asymmetry in
cycles and steadily increases their height and length.

\begin{figure}[h!]
\caption{Time evolution of the scale factor, $a(t)$, of a closed Friedmann
universe with increasing radiation entropy.} %
\includegraphics[width=8cm,height=5cm]{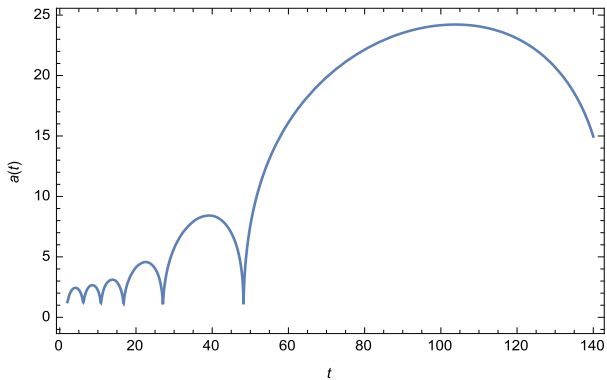}
\end{figure}

All of these insights apply to a universe in which there is no cosmological
constant, $\Lambda $. If we introduce a positive cosmological constant then
there is a big change: the sequence of growing oscillations always ends
(Figure 2). What goes up, need not come down \cite{bdab}. No matter how
small the value of $\left\vert \Lambda \right\vert $, the sequence of
growing oscillations will eventually attain a size that brings the $\Lambda $
term into dynamical play. When that happens, it quickly accelerates the
expansion towards asymptotically de Sitter expansion with a scale factor
that expands exponentially in time. No further expansion maxima occur;
entropy production becomes insignificant; and the dynamics are left close to
flatness and slightly dominated by the energy density in the $\Lambda $, or
`dark energy', field -- a situation not dissimilar to what we observe today 
\cite{planck} although the value of $\Lambda $ remains unexplained.

\begin{figure}[h!]
\caption{Adding a cosmological constant to an oscillating Friedmann
universe with entropy increase always ends the oscillations.} %
\includegraphics[width=8cm,height=5cm]{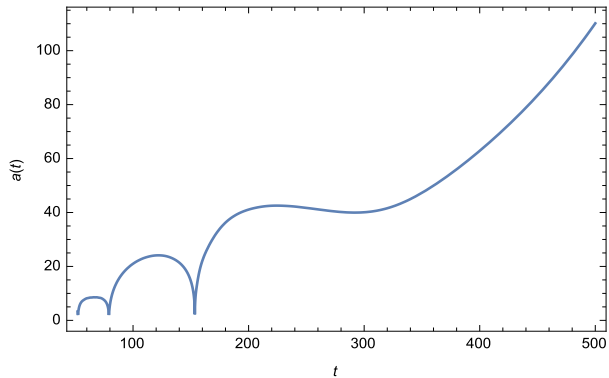}
\end{figure}

This na\"{\i}ve cyclic universe scenario leaves several questions
unanswered. How is it possible to `bounce' through a spacetime singularity
at the end of each cycle with no greater damage to the Friedmann dynamics
than an injection of entropy? And what happens to the shape of more
realistic closed universes over many cycles?

The first difficulty can be ameliorated to some extent by enabling the
bounce to occur at a non-singular finite radius. This can be achieved by
adding a `ghost' field, with negative density, to the Friedmann equations so
that the universe evolves smoothly through a non-singular sequence of finite
oscillations \cite{bkm}. The unitarity of a theory having such a `ghost'
field, has often been called into question. It has recently been shown \cite%
{deRham} that `ghost' fields require the inclusion of operators arising from
a higher energy scale. On imposing shift symmetry, including these operators
allows for a bounce that does not violate tree-level unitarity, at least
within the context of the low-energy effective field theory. If a positive
cosmological constant is included, then the non-singular oscillations must
come to an end and the dynamics evolve towards a future de Sitter state,
just as we described above.

The second difficulty is the one we focus upon. The isotropic Friedmann
universes are not generic solutions of Einstein's equations at late times if
matter is gravitationally attractive and so we need to understand what
happens to closed cyclic universes that are allowed to expand
anisotropically. The most general such universe that allows anisotropic
expansion whilst retaining spatial homogeneity is the Mixmaster universe of
Bianchi type IX, well-studied in connection with its chaotic dynamical
behaviour near a singularity \cite{bkl, jb}. This chaotic behaviour only
occurs on time intervals that include the zero of time and so, if we effect
a bounce at finite expansion radius, we will not need to worry about it. In
fact, if the bounce occurred at the Planck scale, $10^{-43}s,$ then less
than ten oscillations would occur all the way up to the present time of $%
10^{17}s$, because they proceed on a slow logarithmic time scale compared to
the increase in the overall volume of the universe.

The diagonal Mixmaster universe is the most general anisotropic closed
universe with $S^{3}$ spatial topology. Its three orthogonal scale factors, $%
a(t),b(t)$ and $c(t)$, obey the field equation ($8\pi G=c=1$) \cite{LL}

\begin{equation*}
2(\mathrm{ln}\,a)^{\prime \prime \
}+a^{4}-(b^{2}-c^{2})^{2}=a^{2}b^{2}c^{2}\sum_{i=r,g}(\rho _{i}-p_{i}),
\end{equation*}%
and the other two equations are obtained by permuting $a\rightarrow
b\rightarrow c\rightarrow a$. The time derivatives are denoted by $^{\prime
}=d/d\tau \equiv abc$ $d/dt$, where $t$ is the comoving proper time and the
sum on the right-hand side is over the different matter sources present ($r$
= radiation, with $\rho _{r}=3p_{r}\propto (abc)^{-4/3}$and $g$ = ghost
field, with $0>\rho _{g}=p_{g}\propto (abc)^{-2}$). The addition of a
cosmological constant means adding a third $p_{i}=-\rho _{i}$ $=-\Lambda $
constant field to this sum.

When $a=b=c$ these equations describe the isotropic Friedmann metric.
However, when $a,b$ and $c$ are unequal they display remarkably complicated
behaviour because of intrinsically general relativistic effects. Unlike the
simple anisotropic cosmologies which have Newtonian analogues, Mixmaster
universes have anisotropic spatial curvature fashioned by long-wavelength
gravitational waves propagating on an expanding $3$-geometry. Surprisingly,
unlike in an isotropic universe, the sign of the $3$-curvature can also
change with time. When the dynamics are far from isotropy it is actually
negative and so no expansion maximum can occur. Expansion continues until it
is isotropic enough for the $3$-curvature to become positive and only then
can this type of closed universe recollapse, although it still does so
anisotropically.

The Einstein equations for the Mixmaster universe cannot be solved exactly
(in fact they form a non-integrable system) but we can study them
approximately \cite{DLN} and numerically to understand the detailed
evolution over successive cycles if we introduce a `ghost' field. The
`ghost' field produces smooth bounces at finite minima, so that we do not
encounter chaotic Mixmaster oscillations there, and it has no significant
effect on the expansion maxima. For simplicity, we use blackbody radiation
as the matter source (although the results are qualitatively the same if
dust is added) and we can increase the entropy of the universe, as Tolman
envisaged, by injecting new radiation entropy at the start of each cycle.
What happens?

When $\Lambda $ is zero, oscillating Mixmaster universes with increasing
entropy mimic their special Friedmann counterparts in two respects. They
oscillate through a sequence of cycles that grow in size and total lifetime,
and evolve closer and closer to flatness. Yet, here the resemblance with
Tolman's old scenario ends. In each cycle the anisotropy between the three
scale factors grows and is larger at each successive maximum of the volume
(Figures 3a,b). During the subsequent collapse phase of each cycle the
anisotropy continues to amplify.

\begin{figure}[tbp]
\caption{Time evolution of the (a) volume, and (b) individual scale factors, of a Mixmaster universe containing radiation, ghost field and zero $\Lambda$, as it oscillates through many cycles with increasing radiation entropy. The blue dashed, green
dotted and yellow solid lines trace the scale factors $a(t)$, $b(t)$ and $c(t)$. The cycles become increasingly anisotropic.}%
\centering\hfill \break 
\begin{minipage}{0.48\textwidth}
\includegraphics[width=1.0\linewidth, height=0.3\textheight]{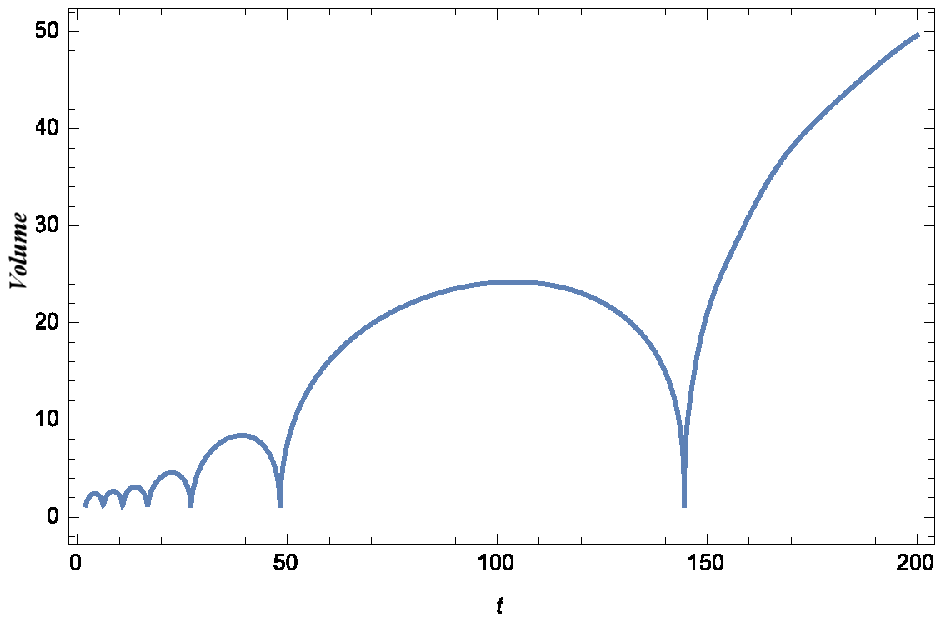}
\subcaption{\label{fig:oscillationswithghostvolume}}
\end{minipage}\hfill 
\begin{minipage}{0.48\textwidth}
\includegraphics[width=1.0\linewidth, height=0.3\textheight]{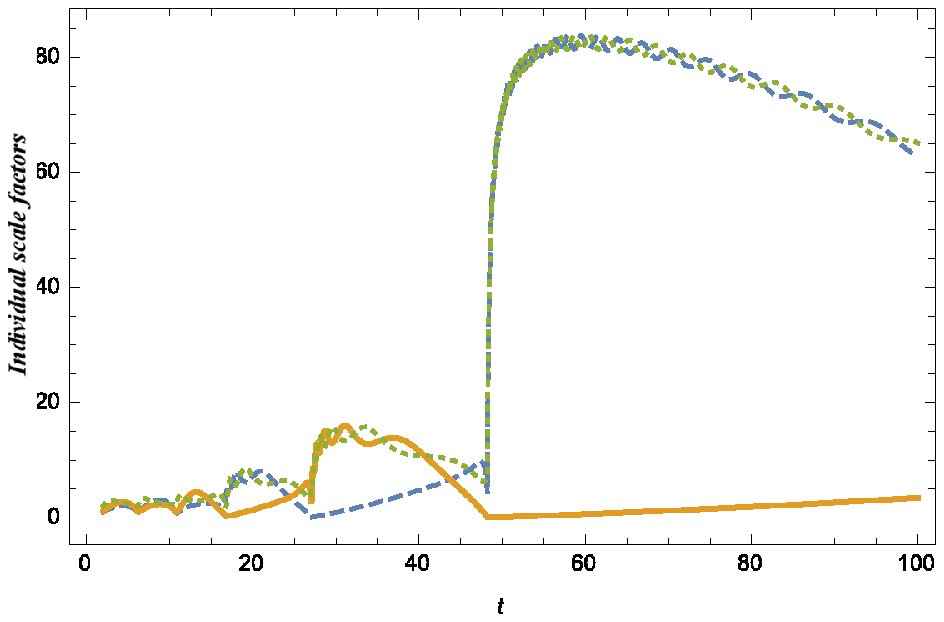}
\subcaption{\label{fig:oscillationswithghostindividual}}
\end{minipage}
\end{figure}

Even if we reduce its cumulative effects by allowing the isotropic `ghost'
field to dominate and produce a bounce at each minimum, the anisotropy level
still grows larger at each successive larger maximum. This type of
anisotropic universe is not like the universe we live in, despite its
proximity to flatness. The appealing features of the isotropic cyclic
universe have disappeared in a chaotic sequence of anisotropic oscillations.

Cyclic universes have recently become popular in theories of gravity that
extend Einstein's general relativity, as can be seen in the extensive review
in reference \cite{brand}. Our findings here will place new pressures upon
these and future scenarios as well. Ekpyrotic models incorporate a
super-stiff field with $p>>\rho >0$ to drive the dynamics to isotropy at the
end of each cycle \cite{ekpy}. But if anisotropic super-stiff pressures
created by collisionless stresses are included in the momentum spectrum then
super-stiff fields will fail to isotropise the collapse \cite{anpr}.
Anisotropies will accumulate and successive Mixmaster expansion maxima will
become increasingly anisotropic. Our own model has various simplifying
features: the metric is diagonal and the fluid flow lines of the radiation
are comoving. We find that dropping these special features does not solve
the growing anisotropy problem. It makes it worse.

\begin{figure}[h!]
\caption{Time evolution of the (a) volume, and (b) the individual Hubble expansion rates, of a Mixmaster universe with positive $\Lambda$, radiation and a ghost field. The
blue dashed, green dotted and solid yellow lines trace the Hubble rates 
$\dot{a}(t)/a(t)$, $\dot{b}(t)/b(t)$ and $\dot{c}(t)/c(t)$. Oscillations cease when $\Lambda$ dominates. The Hubble rates then undergo an anisotropic transition phase before eventually approaching isotropic de Sitter-like expansion where the individual Hubble rates approach the same constant value}%
\centering\hfill \break 
\begin{minipage}{0.48\textwidth}
\includegraphics[width=1.0\linewidth, height=0.3\textheight]{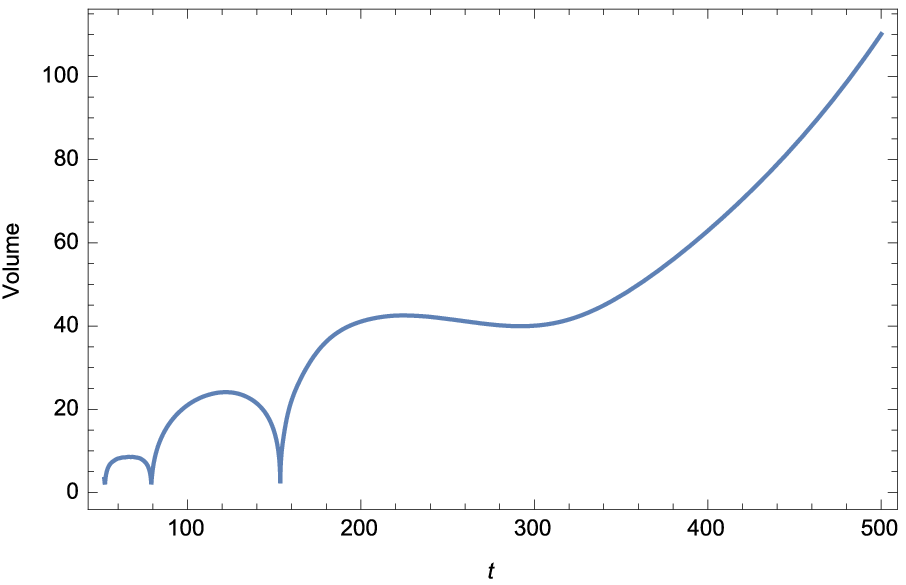}
\subcaption{\label{fig:cosmologicalconstantvolume}}
\end{minipage}\hfill 
\begin{minipage}{0.48\textwidth}
\includegraphics[width=1.2\linewidth, height=0.3\textheight]{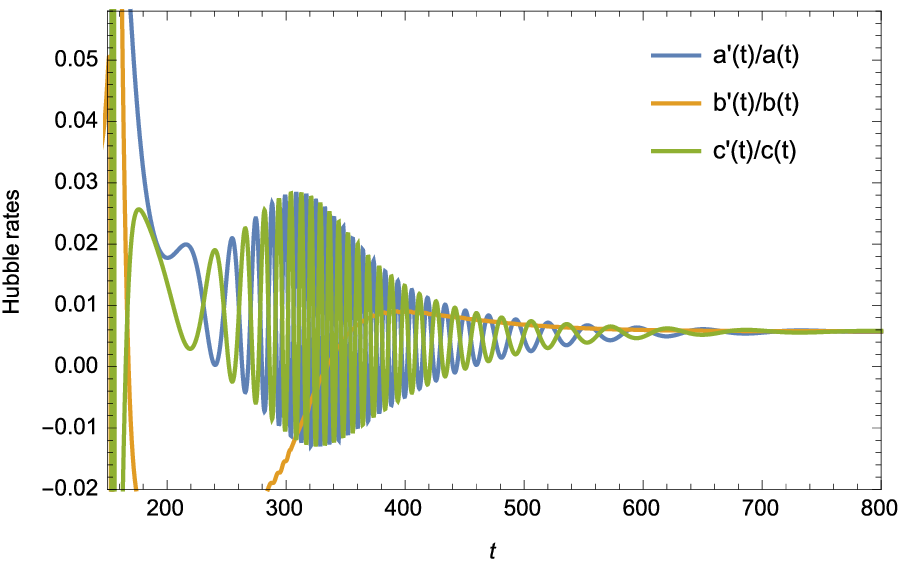}
\subcaption{\label{fig:cosmologicalconstantindividual}}
\end{minipage}
\end{figure}

If we now add a positive cosmological constant to the Mixmaster cycles of a
radiation-filled universe with monotonic entropy increase, the oscillations
still inevitably come to an end, just as in the special Friedmann case --
and for the same reason. Ultimately, the $\Lambda $ stress always dominates
when the cycles grow to a certain size. The death throes of the cyclic
universe yield a distorted period of anisotropic expansion as the dynamics
are accelerated with equal antigravitating force in all directions (Figures
4a,b). We can confirm that the expansion indeed tends to de Sitter as the
Hubble rates in the individual directions tend to a constant value, $\sqrt{%
\Lambda /3}$, when the cosmological constant starts to dominate. The volume
evolves exponentially rapidly towards the asymptotic de Sitter state with $%
a(t)b(t)c(t)\simeq \exp (t\sqrt{3\Lambda })$. If we inhabited the last cycle
where this changeover to $\Lambda $-domination had occurred we would see
evidence of significant anisotropy in the angular intensities of the x-ray
and microwave backgrounds, and probably also in the large-scale Hubble flow,
together with anomalously large abundances of primordial helium-4 as a relic
of the faster anisotropic expansion in the very early stages of the cycle.
This completes our short story of the cyclic universe in general relativity.
\begin{acknowledgments}
JDB is supported by the Science and Technology
Facilities Council (STFC) of the United Kingdom. CG is supported by the
Jawaharlal Nehru Memorial Trust Cambridge International Scholarship.
\end{acknowledgments}

\end{document}